\documentclass[%
 aip,
 amsmath,amssymb,
 reprint,%
]{revtex4-1}

\usepackage{graphicx}
\usepackage{dcolumn}
\usepackage{bm}

\usepackage[utf8]{inputenc}
\usepackage[T1]{fontenc}
\usepackage{mathptmx}
\usepackage{etoolbox}
\usepackage{xcolor}
\usepackage{float}

\makeatletter
\def\@email#1#2{%
 \endgroup
 \patchcmd{\titleblock@produce}
  {\frontmatter@RRAPformat}
  {\frontmatter@RRAPformat{\produce@RRAP{*#1\href{mailto:#2}{#2}}}\frontmatter@RRAPformat}
  {}{}
}%
\makeatother
\begin{document}

\preprint{AIP/123-QED}

\title[]{A cautious user's guide in applying HMMs to physical systems}
\author{M. Schweiger}
\author{A. Saurabh}%
\author{S. Press\'{e}}%
 \email{spresse@asu.edu}
 \homepage{http://www.labpresse.com}
\affiliation{%
Department of Physics, Arizona State University, Tempe, AZ, USA\\
Center for Biological Physics, Arizona State University, Tempe, AZ, USA \\
School of Molecular Sciences, Arizona State University, Tempe, AZ, USA
}%

\date{\today}
\begin{abstract}
Nature, as far as we know, evolves continuously through space and time. Yet the ubiquitous hidden Markov model (HMM)--originally developed for discrete time and space analysis in natural language processing--remains a central tool in interpreting time series data drawn from from physical systems. This raises a fundamental question: What are the implications of applying a discrete-state, discrete-time framework to analyze data generated by a continuously evolving system? Through synthetic data generated using Langevin dynamics in an effective potential, we explore under what circumstances HMMs yield interpretable results.
Our analysis reveals that the discrete-state approximation acts primarily as an abstraction with the inferred states visited in time often more closely reflecting the measurement protocol and modeling choices than features of the underlying physical potential. Crucially, we demonstrate that the states visited over the course of a time series recovered by the HMM can be tuned \textit{a priori} by adjusting the data acquisition scheme even misleadingly recovering reproducible ``intermediate'' states using different HMM tools for a system evolving in a single well potential.
We conclude with a note of measured caution: while HMMs offer a mathematically elegant framework for time series inference, their use in physical modeling should be guided by an awareness of their limitations. In this light, we outline important generalizations of the HMM to continuous space and time and highlight the importance of a well calibrated measurement noise model.
\end{abstract}

\maketitle

\subsection{\label{sec:A} INTRODUCTION: A whirlwind tour of HMMs through the years}

Leonard E. Baum and collaborators introduced the hidden Markov chain in the late 1960's~\cite{baum1966statistical,baum1970maximization} learning probabilistic transition rules governing language, weather patterns, and stock prices. The hidden Markov models (HMMs) that emerged over the subsequent decades~\cite{jelinek1998statistical, rabiner2002tutorial, krogh1994hidden, durbin1998biological, mckinney2006analysis, bronson2009learning} motivated specializations, \textit{e.g.}, Ref.~\cite{qin1997maximum}, and extensions to numerous practical examples across physical applications highlighting at once the versatility and elegance of the HMM's mathematical framework.

In the field of Chemical and Biological Physics, one of the first notable applications of HMMs was to single-molecule fluorescence experiments introduced by Taekjip Ha and co-workers in their studies of DNA Holliday junctions, and RecA protein binding and dissociation~\cite{mckinney2006analysis}. Their work helped establish HMMs as a workhorse for the analysis of noisy time series proving particularly useful across a range of experimental settings especially when supplemented by prior knowledge of the underlying physical system~\cite{mckinney2006analysis, saurabhiii2022single, sgouralis2017introduction, sgouralis2018bayesian, sgouralis2017icon, mor2021systematic}. However, in many cases prior knowledge is incomplete or uncertain, and HMMs are applied in a more agnostic fashion--without clear grounding in the system's physical characteristics--raising questions regarding how best to interpret the HMM analysis output.

In this spirit, we revisit key assumptions--beyond the clear Markov assumption--inherent to  HMMs summarized in Table~\ref{tab:assumptions} and discuss their implications on physical systems. These considerations are especially timely as we continue to push spatial and temporal sensitivity of experimental setups~\cite{mazal2019single,nettels2024single, schanda2024nmr, grabenhorst2024single, foote2024quantifying, phelps2017using, maslov2023sub, otosu2015microsecond, grimm2012high, madsen2021ultrafast, bustamante2021optical}.

As illustrated in the table, consider a system smoothly transitioning through regions of phase space while evolving between deep potential wells which we call here ``basins''. Such smooth transitions, for example in systems like single molecules coupled to bead motion in an optical trap, may be obscured at low data acquisition rates~\cite{wong2006effect, wong2009beyond}. However, as the acquisition rate increases, the transient motion of molecules along the potential can become apparent as shown in Fig.~\ref{fig:epsart} and may be misinterpreted as additional discrete states.

The astute reader may have noticed that the assumption on the number of states is missing from the table. This is because the HMM has already been generalized to learn the number of states. Here, we have in mind nonparametric extensions of the Hidden Markov Model (such as infinite HMMs or iHMMs~\cite{beal2001infinite,sgouralis2018single,sgouralis2017introduction,sgouralis2018bayesian}, or Beta-Bernoulli coupled HMMs~\cite{saurabhi2023single,saurabhii2023single,saurabhiii2022single}) that nonetheless inherit the discrete state assumption of of regular (parametric) HMMs. 

As we will demonstrate, learning the number of states can be a double-edged sword. That is, the number and meaning of the discrete states occupied in the HMM-learned model can be manipulated by reconfiguring the measurement scheme, with no change in the underlying dynamics of the physical system. Below, we examine these issues, offering both caution and optimism regarding the use of HMMs in analyzing physical systems.

\subsection{\label{sec:B}Simulating physical, continuous space, systems analyzed by HMMs}

In practice, treating continuous space and time systems computationally necessitates a discrete time-grid. To approximate continuous time, the grid is chosen sufficiently small in order to describe the differential time-evolution of the physical system undergoing fast, microsecond scale dynamics.

\begin{table}[H]
    \centering
    \begin{tabular}{|p{4.0cm} | p{4.0cm}|}
    \hline
       \textbf{Assumptions:}  & \textbf{Physical implications:}\\
       \hline
       States are discrete  &  No extended regions of phase space should be explored over the course of a transition from one potential energy well--\textit{i.e.}, basin defining an explorable region of phase space--
       to another. Put more succinctly, transitions between states must be infinitely sharp. This also implies that barriers separating potential basins must be both high and narrow to justify the HMM's assumption of discrete, instantaneous switching between states. \\
       \hline
       
       State transitions coincide with the discrete measurement times  &  Changes in regions of phase space occupied by physical systems, \textit{i.e.}, potential basins, must occur infrequently as compared to the measurement timescale. This guarantees minimal rounding error in assuming changes coincide with the measurement time. Perhaps counter-intuitively, when physical changes do occur, however, they must be instantaneous (see previous assumption). \\ 
       \hline
       Waiting times in each state are
        geometrically distributed  & Potential basins in which the system evolves must be sufficiently narrow and allow minimal exploration of phase space over the measurement time intervals. In other words, different basins explored by the system must be well separated in terms of measurement signal output.\\ 
\hline
        Measurements corrupt the instantaneous value of the state attained at each measurement time  & Physical measurements can only be modeled as ``instantaneous'' at each observation time, not ``integrative'' over an observation time window. In doing so we ignore averaging of the  system's behavior over observation windows. As such, if transitions occur at arbitrary (unobserved) times, arbitrarily many transitions cannot be inferred \emph{between} or \emph{during} measurements, making it impossible to learn the system's state at any time.\\
        \hline
    \end{tabular}
    \caption{\textbf{Key HMM assumptions and their physical implications.}}
    \label{tab:assumptions}
\end{table}
Throughout, we will simulate a continuum system by the Brownian motion of a single degree of freedom $x$ (representing, for example, a reaction coordinate, or the sum of a large number of degrees of freedom) in an effective potential. We will use the overdamped Langevin equation as a minimal model for Brownian motion in a potential $U(x)$ with friction (of coefficient $\zeta$) and thermal fluctuations $\langle r(t),r(t') \rangle = \delta(t-t')$ capturing continuous time and space evolution  
\begin{equation}
    \zeta \frac{d x}{dt} = -\nabla U\left(x\right) + r\left(t\right),
\end{equation}
though more general formulations of the dynamics, \textit{e.g.}, generalized Langevin, are also possible. 
Integrating the Langevin equation under the It\^{o} approximation yields
\begin{equation}
    X_{n+1} | x_n \sim \mathbf{Normal} \left(x_n + \frac{\Delta t}{\zeta} f\left(x_n\right), \frac{2 \Delta t k T }{\zeta}\right),
\end{equation}
forming a trajectory over time with $\mathbf{Normal}$ understood as the Normal distribution. The expression above reads, ``the random variable $X_{n+1}$ conditioned on the realization of the previous random variable $x_n$ is drawn from a Normal distribution''.
As per convention, we use the uppercase $X_{n}$ to denote the random variable (which attains no particular value until it is realized), and the lowercase $x_{n}$ to denote a particular realization (the value that $X_{n}$ attains upon its realization). Adhering to the notion of measurement of the HMM paradigm for more direct later comparison, we assume that the measurement output, $w_{n}$, depends on the instantaneous $x_{n}$ and potentially additional parameters $\sigma$ of the measurement system as follows
\begin{equation}
    W_n | x_n, \sigma \sim \mathbb{G}\left(x_n,\sigma\right).
\end{equation}
The distribution $\mathbb{G}$ above is commonly referred to as the ``emission'', ``noise'', or ``measurement'' distribution. 

To emulate typical Normal noise detection models, we will make the choice here that
\begin{equation}
    W_n | x_n, \sigma_{read} \sim \mathbf{Normal}\left(x_n,\sigma_{read}\right)
\end{equation}
assuming that measurements corrupt the instantaneous value of the state attained at each measurement time. More complex generalizations involving integrative detector models are explored elsewhere~\cite{kilic2021generalizing} and not critical to the arguments we will make shortly. In this paper, to allow us to directly compare the system's positions $x$ to noisy measurements ($w$), readout noise is simulated in the same units as the dynamics ($\text{nm}$). Furthermore, later in the paper, to imitate real experiments, we will modify the Normal emission distribution above by first taking box-car averages of the trajectories over multiple time-steps and then applying the stochastic noise to reduce the temporal resolution of the simulation.

\begin{figure*}
\includegraphics[width =\textwidth]{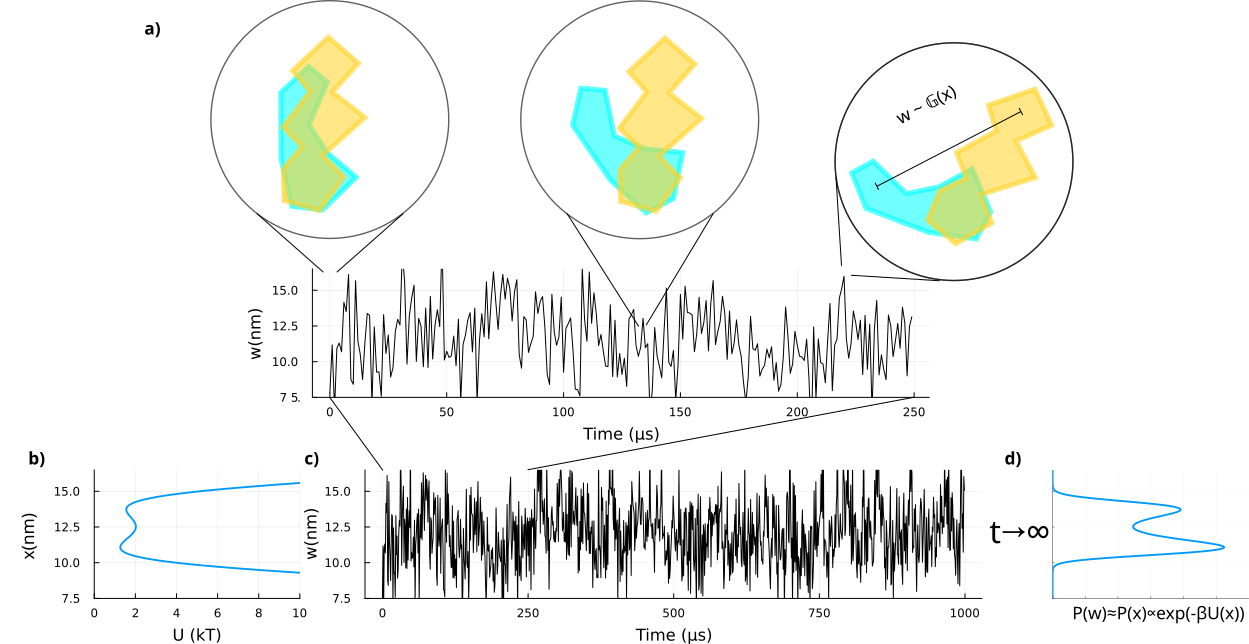}
\caption{\label{fig:epsart} 
\textbf{An example potential in continuous space and its output. a)} A cartoon of a system's (\textit{e.g.}, a protein's) instantaneous configurations in time. 
\textbf{b)} The 1D potential in which the system evolves, in units of $kT$ with $k$ the Boltzmann's constant and $T$ the temperature, versus the position in units of nanometers. \textbf{c)} The resulting time series with a zoomed in region right above. We highlight the two layers of stochasticity to consider in this time series: 1) the position itself is stochastic due to the thermal fluctuations as the particle evolves in the potential of panel \textbf{b)}; and 2) measurement noise corrupts the instantaneous position into a readout $w$ in this trajectory simulated to replicate a constant-force optical tweezer experiment~\cite{bustamante2021optical, grimm2012high}. \textbf{d)} Ignoring measurement noise (\textit{i.e.}, assuming the emission distribution $P(w)$ approximates the position distribution $P(x)$), the accumulation of many observations leads to a distribution approaching the equilibrium (Boltzmann) distribution associated with the potential:
$P(x) \propto \exp\left(-\beta U(x)\right)$.
In practice, however, data are finite and convolved with measurement noise, making, as we will see in more detail, simple histogramming a flawed method for inferring the number of basins.
}
\end{figure*}

As such, the time series we simulate here depend on two sources of noise: thermal fluctuations derived from the Langevin dynamics and the measurement noise; see Fig.~\ref{fig:epsart}c.

\subsection{\label{sec:C}Breaking the HMM}

We are now in a position to investigate the HMM's analysis output by simulating Langevin dynamics of a system trapped either a single well or double well (smooth) potentials as we vary the: 1) measurement noise level; and 2) for double well potentials, the barrier width and height both relevant as compared to the variance of the thermal fluctuations and measurement noise. 

\begin{figure*}
\includegraphics[width = \textwidth]{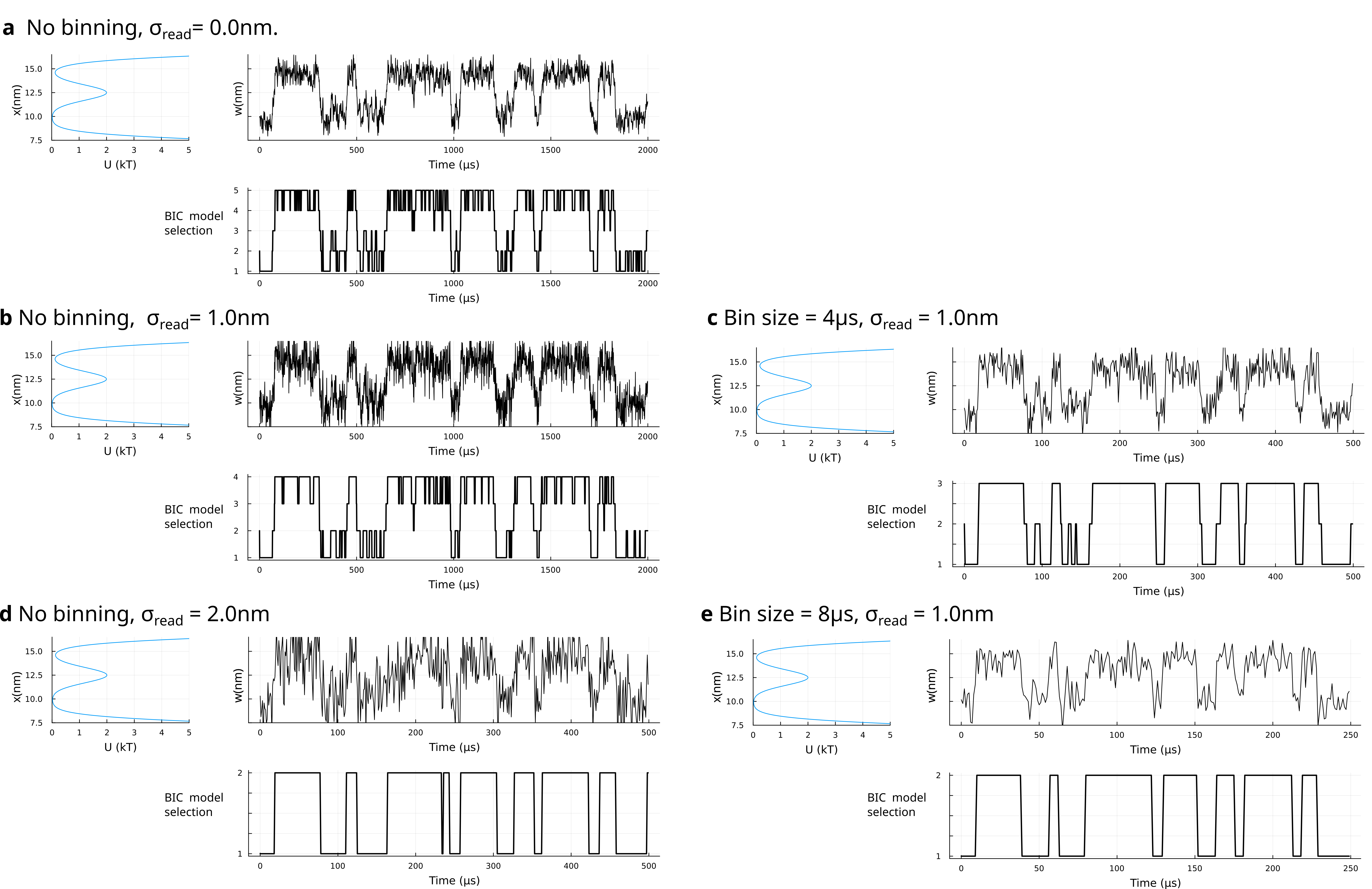}
\caption{\label{fig:bic} \textbf{HMM output for data generated from a system evolving in a double well potential under various measurement noise levels and data acquisition rates}. 
\textbf{a.} On the left, a double well potential with barrier height of approximately 2~kT. On the right, trajectory of a system undergoing Langevin dynamics under this potential at a temperature of 300~K dampened by a friction coefficient of 0.0002 g/s. A time step size of $1.0~\mu$s is chosen such that the fluctuations along the potential are well-sampled as determined by the shape of the potential near the minima. On the bottom, a discrete state trajectory predicted using an HMM (with a BIC criterion implemented, as described in the text, to determine the number of states). \textbf{b-e.} Same as in \textbf{a} but the trajectory is binned and readout noise is applied. As we will discuss in the main body, counter-intuitively, we notice the correct number of basins (``states'') being identified when measurement noise is greater or as we reduce acquisition rates where smooth transition between basins are ``averaged out''.}
\end{figure*}

Immediately, we notice a breakdown for a system evolving in a double well potential when analyzing the time series with the HMM under a range of assumed state counts, and applying the commonly used Bayesian information criterion~(BIC)~\cite{schwarz1978estimating} to select the optimal state number. The BIC is just one example of a range of strategies using a regularized model criterion to learn the number of states, including the AIC~\cite{akaike1998information}, likelihood ratio test~\cite{pishro2014introduction}, or even Bayes Factor~\cite{morey2016philosophy}), which may all yield slightly different results. 

Interestingly, the ``best case scenario''--the limit of high data acquisition rate and no measurement noise with only thermal noise--shown in  (Fig.~\ref{fig:bic} a) exhibits one of the worst breakdowns of HMM analysis.
The analysis here identifies a 5-state model (the most complex model explored). The ``extra'' states appearing in this analysis originate from the smooth transitions between basins (\textit{i.e.}, the emergent temporal correlations in overdamped Langevin dynamics).

To explore the hypothesis that the extra states originate from smooth transitions, we can either ``bin'' the data in time (with a boxcar-average as we did in Fig.~\ref{fig:bic} b,d) or progressively increase the measurement noise ($\sigma_\text{read}$) as we did in Fig.~\ref{fig:bic} c,e. 

Both ``work''. That is, as the data acquisition rate deteriorates or the measurement gathers larger error, eventually the HMM analysis supplemented by the BIC identifies two basins. Both processes ultimately ``average out'' motion over the barrier that could be picked up as additional states.

\begin{figure*}
\includegraphics[ width = \textwidth]{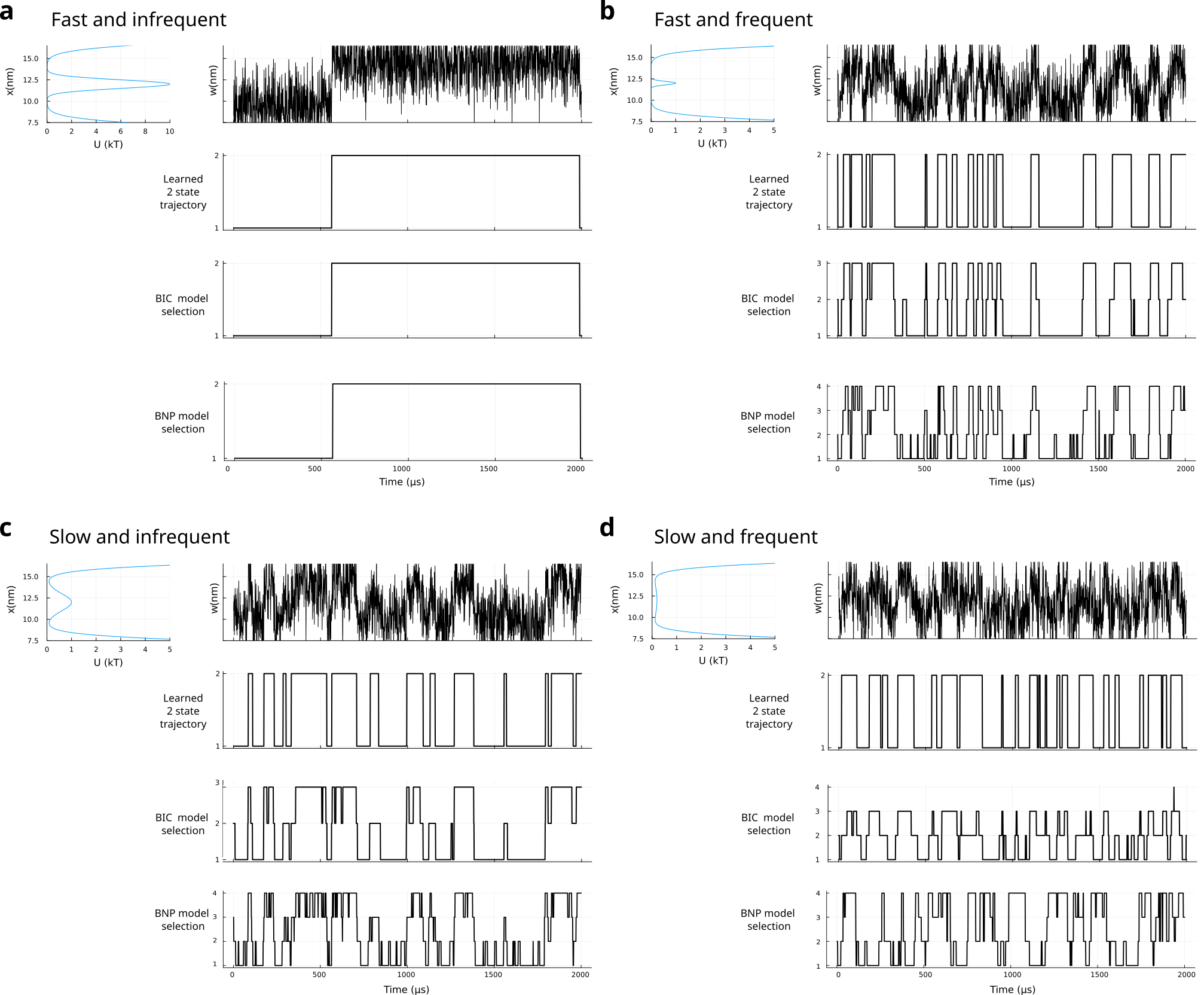}
\caption{\label{fig:2st} 
\textbf{HMM predictions for systems undergoing Langevin dynamics in double-well potentials.} All trajectories are generated using the same temperature, friction, coefficient, and time-step size as in Fig.~2 with no binning and no measurement noise. \textbf{a.} High barrier height with narrow wells lead to fast thermal fluctuations but very infrequent barrier crossings. Below the simulated data on the right, a fixed 2-state (parametric) HMM is used to analyze the simulated data with estimated trajectory closely matching the signal. Right below, a BIC corrected HMM is used to analyze the same simulated data while, in the bottom panel, a Bayesian nonparametric (BNP) HMM is used.  \textbf{b.} As the barrier is lowered, crossings become more frequent and states become less distinct. The nonparametric now predicts different state numbers and trajectories from those obtained using parametric HMM and BIC corrected HMM analysis. \textbf{c.} If the barrier is high but widened, the transitions occur more slowly/continuously. Inconsistencies between analyses appear as in b. \textbf{d.} Barrier height is lowered further. At this point, is it reasonable to ask: are the basins identifiable from the signal? This challenging situation is evident in both nonparametric and BIC corrected fits where even higher dimensional 4-state models are predicted.}
\end{figure*}

Next, in Fig. ~\ref{fig:2st} we investigate how features of the potential under the ``best case scenario'' of Fig.~\ref{fig:bic}a impact HMM analysis as we vary features of the potential, such as the barrier width and height. Indeed, whereas HMM models use  transition probabilities to describe evolution between potential basins, Langevin models allow the barriers between states to vary in height and vary in width independently. In doing so, we can simulate both ``fast'' and ``slow'' transitions and tune transition frequencies (``frequent'' vs ``infrequent''), in a manner that is difficult for HMMs to capture. As we will see, when data are acquired at a similar time-scale to the Langevin dynamics correlations, phase space explored about the basins can appear as additional states, the number of which can be tuned by binning the data. 

To illustrate this, we present the analysis for a simple 2-state HMM with no BIC regularization, BIC corrected HMM (as in Fig.~\ref{fig:bic}), and a more rigorous Bayesian nonparametric (BNP) HMM based on a Beta-Bernoulli process, which we favor over the iHMM's hierarchical Dirichlet process for its intuitive clarity~\cite{saurabhi2023single,saurabhii2023single,saurabhiii2022single}.

As we notice immediately, while the 2-state HMM works ``by definition'', both frequent or slow transitions return results inconsistent with the number of anticipated basins for the BIC corrected HMM and BNP HMM. The cutoff in signal for high signal is only a matter of plotting a finite range of the signal.

More interesting still is the case where the potential has a unique, even flat, basin, simulated once again under the ``best case scenario'' of Fig.~\ref{fig:bic}a (no binning and no measurement noise). Here we analyzed the data once more using a 2-state and 3-state HMM with no BIC regularization, BIC corrected HMM, and a BNP HMM. Importantly, and deeply misleadingly by eye, the data appears to visit multiple states. Compounding the difficulty, the 2-state HMM returns what it was meant to output: a clean 2-state trajectory visiting entirely fictitious states for this system evolving in a single well potential. The BIC corrected HMM and BNP HMM also do as intended: they interpolate the smooth dynamics by positing possible (``intermediate'') states. Adding to a false sense of security: the 3-state model trajectory learned is essentially identical with the BIC corrected HMM which identifies 3 states (a candidate short-lived intermediate state) for a system evolving in a single well. None of the basins identified, whether by the BNP HMM or any other method, exist.

\begin{figure}
\includegraphics[ width = 0.5\textwidth]{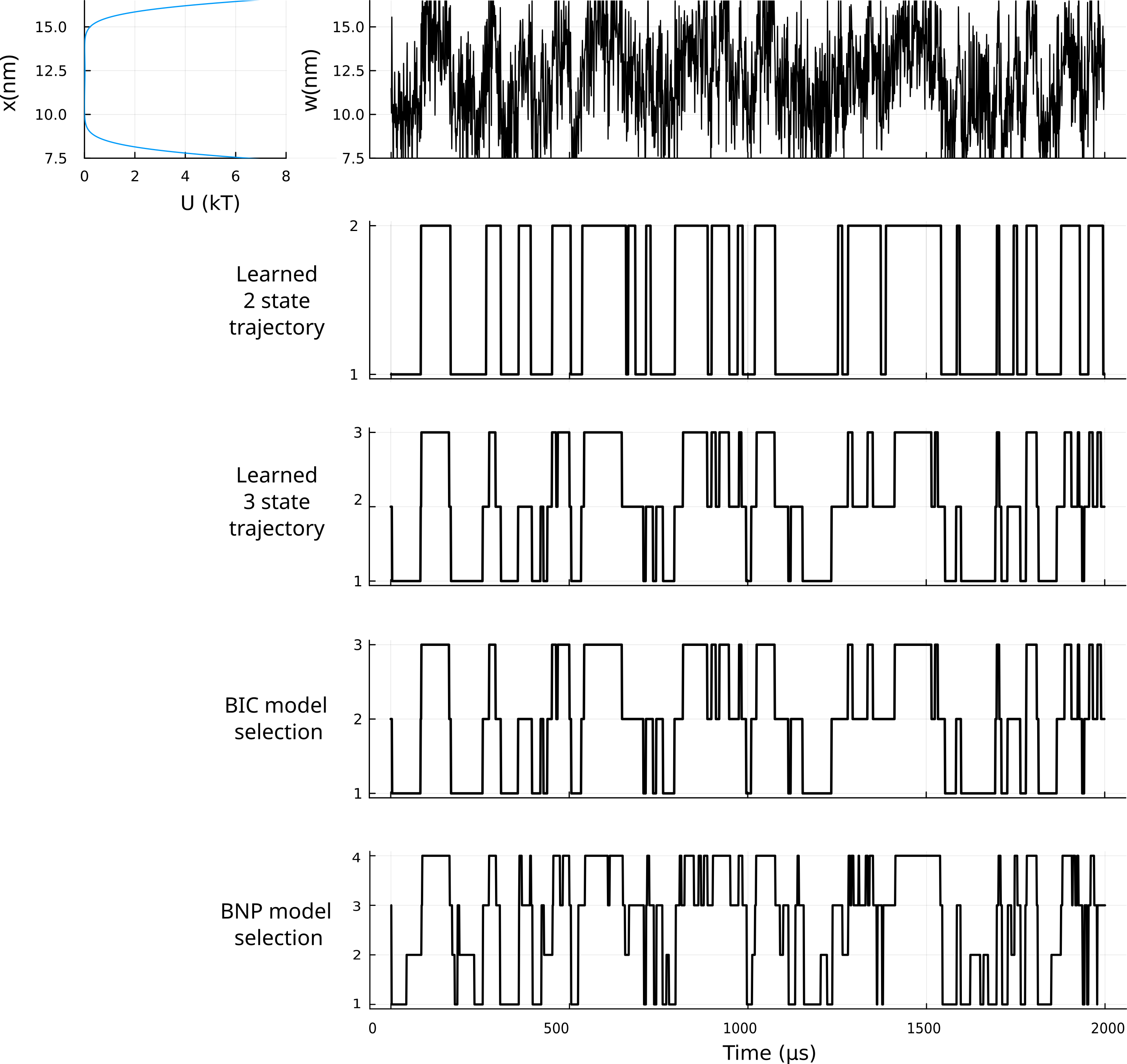}
\caption{\label{fig:1st} \textbf{HMM predictions for a system undergoing Langevin dynamics in a single well potential}. The system trajectory is simulated using the same values for temperature, friction coefficient, and time step size as in Fig.~2 with no binning or additional of noise. As done in Fig.~\ref{fig:2st}, we estimated trajectories using a 2-state parametric model~(second row), a 3-state parametric model~(third row), BIC corrected HMM~(fourth row), and a BNP HMM~(bottom row). }
\end{figure}

\subsection{\label{sec:D}A first discussion by way of an intermission}

Many of the procedures we have invoked in Fig.~\ref{fig:bic} have allowed us to converge to interpretable results. That is, data binning and added noise have helped us match the number of HMM states to the expected number of basins. Why is this not a convincing solution to our problem?

Of course, we had the advantage of knowing the number of basins, \textit{i.e.}, ground truth. When ground truth is unknown, often justifying an experiment's \textit{raison d'\^{e}tre} to begin with, we could have chosen to bin our way down to the desired state number.  

Similarly, whether by adding noise, as explored in Fig.~\ref{fig:bic}, or using highly idealized (\textit{c.f.}, incorrect) emission distributions as explored elsewhere~\cite{saurabh2024statistical} we can cheat the system and obtain the state number desired.

Naturally, these solutions throw rigor the wind: Binning data undermines the time resolution of the detector and only increase the complexity of the emission distribution when modeled correctly; in keeping with this logic, there is no known theoretical justification for corrupting, \textit{i.e.}, adding noise often drawn from statistics differing from those in the original data, to existing data. Likewise caution should be exercised when introducing approximations, often left unverified, when invoking a modular BIC regularization separating kinetic parameter and state number determination as two disjoint steps. This greedy approach is in contrast to a joint state number and kinetic parameter inference afforded by Bayesian nonparametrics that nonetheless still inherit the discrete state approximation. 

We now take a step back if only to say that, had we not been motivated by the elegance of the HMM's framework, our guide would abruptly conclude with the following terminal diagnostic: the HMM, as applied to physical systems, yields the answer desired prior to collecting data in the absence of orthogonal experimental results constraining the analysis. 

In the second half of this work, we take a distinctly mathematical tone to circumvent \textit{ad hoc} patch ups such as binning and redress the challenges we've presented. The reader can now choose to skip the next section as desired and jump directly to the discussion.

\subsection{\label{sec:E}HMMs: A mathematical look under the hood}

Before turning to HMM generalizations, we provide a more complete mathematical description of the HMM.

HMMs describe the states $S_0, S_1, S_2,\dots$ of a Markov chain at discrete times $t_0, t_1, t_2 ,\dots$, evolving within a discrete state space, such that 
$S_n \in \{\sigma_{m}\}_{m=1}^M$ for all $n$, with M denoting the state space size. For simplicity, we denote $\{\sigma_{m}\}_{m=1}^M$ by $\sigma_{1:M}$.
Given an initial probability vector $\vec{\rho} = [P(s_0 = \sigma_0),P(s_0 = \sigma_1),\dots,P(s_0 = \sigma_M)]$ over the set of states $\sigma_{m}$, the system evolves according to the transition model
\begin{equation}
    S_{n+1} | s_{n} \sim \mathbf{Categorical}(\vec{\pi}_{s_n}),
\end{equation}
where $\mathbf{Categorical}$ distribution is the generalization of the Bernoulli distribution for more than two possible outcomes. Here, $\vec{\pi}_{s_n} = [\pi_{s_{n} \rightarrow \sigma_{1}}, \pi_{s_{n} \rightarrow \sigma_{2}}, \dots, \pi_{s_{n} \rightarrow \sigma_{M}}]$ is a vector of transition probabilities whose elements are normalized to unity. Each element, $\pi_{s_{n} \rightarrow \sigma_{m}}$, denotes the probability of a transition from state $s_{n}$ to state $\sigma_m$ occurring between the measurements taken at $t_{n}$ and $t_{n+1}$. An emission model captured by $\mathbb{G}$ ($W_{n}|s_n \sim \mathbb{G}(s_n,\sigma)$), completes this forward model.
While often, in the literature, we generically write the emission distribution as
\begin{equation*}
    \mathbb{G}(w_{n};s_n,\sigma) = \mathbf{Normal}(w_{n}|s_n,\sigma_{read}),
\end{equation*}
this emission model can also be adapted to different experimental setups. For example, in F\"{o}rster resonance energy transfer (FRET) experiments, an EMCCD camera may be used to observe biomolecules, where an additional electron multiplication stage must be modeled in addition to Gaussian readout and photon shot noise~\cite{hirsch2013stochastic}.

By itself, both the transition and emission models are sufficient to write down a data likelihood for the HMM for all data $w_{n}$ ranging from $1:N$ denoted $w_{1:N}$, namely $\mathbb{P}(w_{1:N}|\theta)$ with $\theta$ collecting all unknown parameters. These parameters include parameters from the transition or emission models such as the probability of transition from any state to any other state as well as the emission model variance, for instance. As written, the likelihood is understood as the probability of the data given (or ``conditioned on'') the parameters, $\theta$. Maximizing the likelihood, whether for the HMM or any other application is then equivalent to finding the parameters making the sequence of data as observed most probable.  

A slightly more informative probability, $\mathbb{P}(\theta|w_{1:N})$ understood as how probable the parameters are given the observed data is called a posterior. To construct this object, we need to supplement the transition and emission models with priors 
~\cite{mckinney2006analysis, saurabhiii2022single, sgouralis2017introduction, sgouralis2018bayesian, sgouralis2017icon, mor2021systematic}.
Under this more general scenario, we write 
\begin{align}
    \vec{\rho} &\sim \mathbf{Dirichlet}(\vec{\alpha}) \\
    \vec{\pi}_{\sigma_n} &\sim \mathbf{Dirichlet}(\vec{\alpha}) \\
    S_0|\vec{\rho} &\sim \mathbf{Categorical}\left(\vec{\rho}\right) \\
    S_{n+1}|s_{n} &\sim \mathbf{Categorical}\left(\vec{\pi}_{s_n}\right)\\
    W_{n}|s_n &\sim \mathbb{G}(s_n,\sigma),
\end{align}
where the Dirichlet priors, with hyperparameters $\vec{\alpha}$, are discussed in greater detail elsewhere~\cite{presse2023data}.

\subsection{\label{sec:F}Relaxing the HMM assumptions}

\subsubsection{\label{sec:1} Continuous time}

The continuous time alternative to the HMM which, while still assuming discrete states allows state transitions to occur at any point between instantaneous measurements, can be formulated by re-writing transition probabilities using 
rates $\lambda_{\sigma_m \rightarrow \sigma_{m'}}$. 
In this formulation, the duration of time spent in $s_n$ (called the holding time $h_{n}$) is stochastic and, under the Markov assumption, is drawn from an exponential distribution~\cite{gillespie1977exact, presse2023data}
\begin{equation}
    H_n \sim \mathbf{Exponential}\left(\sum_{m}'(\lambda_{S_n \rightarrow \sigma_m})\right).
\end{equation}
This probability for the waiting time is derived assuming $s_n$ can escape to any of the states 
$\sigma_m$ excluding the state occupied at $s_n$ (and thereby 
requiring the restricted summation). 

By conservation of probability, the transition rule reads as follows~\cite{gillespie1977exact}
\begin{align}
    &S_{n+1} | s_{n} \sim \\ &\mathbf{Categorical}\left(\frac{[\lambda_{s_{n} \rightarrow \sigma_{1}, }, \lambda_{s_{n} \rightarrow \sigma_{2}, }, \dots, \lambda_{s_{n} \rightarrow \sigma_{M} }]}{\sum_{m}'\left( \lambda_{s_{n} \rightarrow \sigma_m} \right)}\right).
\end{align}

Taken together, the sequence of states $\mathbf{S} = [s_0, s_1, s_2, \dots]$ and holding times $[h_0, h_1,\dots]$ completely describe the system's trajectory over time.

With a complete forward model at hand, 
\begin{align}
    \vec{\rho} &\sim \mathbf{Dirichlet}(\vec{\alpha}) \\
    \vec{\lambda}_{\sigma_m \rightarrow \sigma_{m'}} &\sim \mathbf{Gamma}(\alpha, \beta) \\
    S_0|\vec{\rho} &\sim \mathbf{Categorical}\left(\vec{\rho}\right) \\
    &S_{n+1} | s_{n} \sim \\ &\mathbf{Categorical}\left(\frac{[\lambda_{s_{n} \rightarrow \sigma_{1}, }, \lambda_{s_{n} \rightarrow \sigma_{2}, }, \dots, \lambda_{s_{n} \rightarrow \sigma_{M} }]}{\sum_{m}'\left( \lambda_{s_{n} \rightarrow \sigma_m} \right)}\right)\\
        H_n &\sim \mathbf{Exponential}\left(\sum_{m}'(\lambda_{S_n \rightarrow \sigma_m})\right)\\
    W_{n}|s_n &\sim \mathbb{G}(s_n,\sigma),
\end{align}
all that remains is the mathematical exercise of inverting the model to infer parameters of interest~\cite{kilic2021generalizing} in a manner analogous to the HMM.

This model relaxes the assumption that transitions occur instantaneously at measurement times thought it retains the assumption that reactions are instantaneous. That is, the Hidden Markov Jump Process (HMJP, the name for the continuous time relaxation of the HMM's assumptions) still models \emph{discrete space}, ignoring the smoothness of the potential inherent to Langevin dynamics and ensuing correlations in the data. Thus, even as a matter of principle, the HMJP does not address the issues identified in ~\ref{sec:C} though it helps in modeling dynamics between narrow basins separated by narrow barriers not as high and potentially exceeding the measurement timescale ~\cite{kilic2021generalizing, kilic2021extraction}.

\subsubsection{\label{sec:2}Continuous space}

Langevin dynamics describing a system's evolution in a potential anticipates measurements with time correlations due to the smoothness of phase space. These expectations fall outside the purview of typical discrete state HMM approximations and
decorrelated HMM emissions from previously occupied states. 

To model continuous space and curtail these difficulties, we consider placing a general prior $\mathbb{P}(\cdot)$ over the curve $U(x)$. The forward model now reads
\begin{align} 
    U(x) &\sim \mathbb{P}(\cdot)\\
    X_{n+1} | x_n &\sim \mathbf{Normal} \left(x_n + \frac{\Delta t}{\zeta} f\left(x_n\right), \frac{2 \Delta t k_B T }{\zeta}\right) \\
    W_n | x_n, \sigma &\sim \mathbb{G}\left(x_n,\sigma\right),
\end{align}
where we have dropped the problem-dependent initial condition $P(X_0)$ for simplicity. As before, statistical inversion of this generative model is well defined~\cite{bryan2020inferring} and can be generalized to higher dimension by substituting the single dimensional Normal distribution for a multivariate Normal. 
Concretely, the Gaussian Process ($\mathbf{GP}$) may serve as a prior over all candidate curves~\cite{williams1995gaussian,bryan2020inferring}
\begin{equation}
    U(x) \sim \mathbf{GP}(\kappa(\cdot,\cdot))
\end{equation}
with kernel function $\kappa(\cdot,\cdot)$ essential in promoting smoothness and thus spatial correlations \emph{a priori} in the potential $U(x)$.

\subsection{\label{sec:G}DISCUSSION: B\lowercase{eyond the} HMM}

In some sense, this section is a follow up on the primary discussion of Sec.~\ref{sec:D} in using HMMs for analyzing data drawn from physical systems.

Despite that section's admonitions, HMMs are an elegant, rigorous, and internally consistent set of mathematics when interpreted within their own assumptions. Indeed, once we have accepted the assumptions inherent to the HMM framework, then computing quantities such as the optimal (Viterbi) path are foregone, deterministic, conclusions reassuringly leaving no mathematical wiggle room for improvisation. 

Yet trying to lift the assumptions inherent to HMMs for physical systems often presents unique challenges. Take, for instance, a continuous space model with an emission distribution modeling the integration of output by a detector over a finite exposure window: analysis will struggle as there exist infinite-fold degeneracies over a manifold of trajectories that may be followed over an integrated exposure. 

Further absent from our discussion thus far is also the notion of whether a single reaction coordinate is sufficient in describing the system's evolution in phase space and the possibility of learning about ``degenerate'' states with the same emission level distinguishable by their kinetics alone~\cite{gotz2022blind, agam2023reliability}. 

Beyond physical considerations, computational considerations as well have motivated the substitution/enhancement of many classical numerical analysis frameworks leveraging artificial intelligence (AI) tools~\cite{asadiatouei2024deep, duncker2019learning}. AI tools have been used to improve cost: Asadiatouei \textit{et al.}~\cite{asadiatouei2024deep} used deep learning-based method to detect abrupt changes in time series at lower computational cost than classical HMM step detection, especially in noisy, high-throughput, scenarios. To generalize models:
Duncker \textit{et al.}~\cite{duncker2019learning} relaxed one of the assumptions identified here, approximating stochastic dynamics with an interpretable continuous-time latent variable neural model, already applied to biological systems to learn models in continuous time, analogous to our proposed continuous-space models of Langevin dynamics. To introduce original models of similar transition kinetics: Krishnan \textit{et al.}~\cite{krishnan2017structured} introduce structured inference networks to approximate a model of latent dynamics similar to HMMs. To ease the pain of long convergence times of traditional Monte Carlo methods: Rudner \text{et al.}~\cite{rudner2022learning}, frame stochastic dynamical systems as differentiable latent models optimized by gradient descent.

In fact, as AI and HMMs developed contemporaneously, there is a long history of joint efforts  ~\cite{niles1990combining}. Modern approaches build on this foundation by integrating deep neural networks into HMM architectures. For instance, Tran \textit{et al.}~\cite{tran2015variational} used variational autoencoders for inference in time-series models with HMM structure. More recently, Bansal \textit{et al.}~\cite{bansal2024hmm} used neural network models of emissions and transitions, enabling expressive non-Markovian dynamics to emerge from training data. Hybrid models like these preserve the latent states' interpretability, while exploiting deep learning's representational power.

These extensions, such as expressive and trainable emission models, are not only exciting but also underscore perhaps the most critical point of all. Correctly calibrated emission models can help mitigate data misinterpretation. Put differently, whether the dynamics are discrete or continuous in space, it is ultimately physics-informed and well-calibrated emission models that probabilistically determine the degree to which a change in successive points is attributable to measurement noise versus the evolution of the underlying physical system. A carefully calibrated emission model is therefore critical. 

As such, the notion that there should exist a ``validated'' or ``yet to be validated'' HMM package broadly applicable to all data from single-molecule force spectroscopy or all FRET data, is perplexing. Even more confounding is the notion that the perceived validity of a package should depend on its widespread use. In the same vein, the belief that analysis packages can be blindly compared and bench-marked on common datasets only further deepens scientific concern. In reality, what exists are two key notions: 1) mathematical frameworks (distinct from specialized packages) that help organize our thinking; and 2) the need for careful calibration of noise models specific to each scientist's experimental setup--an essential first step in any quantitative analysis.

\subsection{Data Availability}

All data and code supporting the findings of this study will be made publicly available upon acceptance/publication. They can also be shared earlier upon reasonable request to the corresponding author.

\begin{acknowledgments}
SP acknowledges support from the NIH (R01GM134426, R01GM130745, RF1MH128867, and R35GM148237), and US Army (ARO W911NF-23-1-0304).
\end{acknowledgments}

\nocite{*}
\bibliography{bibliography}

\end{document}